%% file: SPP_Reversal.tex
\documentclass[12pt]{iopart}

\usepackage{iopams,setstack,graphicx,here}
\usepackage{cite}
\eqnobysec

\usepackage[usenames,dvipsnames]{color}

\newcommand{\mean}[1]{\left < #1 \right >}
\newcommand{\abs}[1]{\left | #1 \right |}
\renewcommand{\vec}[1]{\bi{ #1 }}

\newcommand{\eqnlabel}[1]{\refstepcounter{equation}\label{#1}\addtocounter{equation}{-1}}
\newcommand{\eqref}[1]{(\ref{#1})}

\newcounter{Aeqnval}
\def\Anumparts{\addtocounter{equation}{1}%
     \setcounter{Aeqnval}{\value{equation}}%
     \setcounter{equation}{0}%
     \def\theequation{\ifnumbysec
     \Alph{section}.\arabic{Aeqnval}{\it\alph{equation}}%
     \else\arabic{Aeqnval}{\it\alph{equation}}\fi}}

\def\endAnumparts{\def\theequation{\ifnumbysec
     \Alph{section}.\arabic{equation}\else
     \arabic{equation}\fi}%
     \setcounter{equation}{\value{Aeqnval}}}

\usepackage{pdfcomment}

\makeatletter
\newcommand{\annotate}[2][]{%
\pdfstringdef\x@title{#1}%
\edef\r{\string\r}%
\pdfstringdef\x@contents{#2}%
\pdfannot
width 2\baselineskip
height 2\baselineskip
depth 0pt
{
/Subtype /Text
/T (\x@title)
/Contents (\x@contents)
}%
}
\makeatother

\begin{document}


\title[Diffusion properties of active particles with directional reversal]{Diffusion properties of active 
particles with directional reversal}
\author{R Gro{\ss}mann$^1$, F Peruani$^2$ and M B\"ar$^{1}$}

\address{$^1$ Physikalisch-Technische Bundesanstalt Berlin, Abbestr. 2-12, 10587 Berlin, Germany}
\address{$^2$ Laboratoire J. A. Dieudonn\'{e}, Universit\'{e} de Nice Sophia Antipolis, UMR 7351 CNRS, Parc 
Valrose, F-06108 Nice Cedex 02, France}

\ead{grossmann@physik.hu-berlin.de}

\begin{abstract}	
The diffusion properties of self-propelled particles which move at constant speed and, in addition, reverse 
their direction of motion repeatedly are investigated. 
The internal dynamics of particles triggering these reversal processes is modeled by a stochastic clock. 
The velocity correlation function as well as the mean squared displacement is investigated and, furthermore, 
a general expression for the diffusion coefficient for self-propelled particles with directional reversal is 
derived. 
Our analysis reveals the existence of an optimal, finite rotational noise amplitude which
maximizes the diffusion coefficient. 
We comment on the relevance of these results with regard to biological systems and suggest further 
experiments in this context. 
\end{abstract}	

%
%

\submitto{\NJP}

\maketitle


\section{Introduction}

Active matter systems~--~ensembles of self-driven particles~--~are a central subject of nonequilibrium 
statistical physics~\cite{toner_hydrodynamics_2005,romanczuk_active_2012,marchetti_hydrodynamics_2013,
menzel_tuned_2015}:~examples include micron-sized active colloids and rods driven by chemical 
reactions~\cite{paxton_catalytic_2004,palacci_colloidal_2010} or by the Quincke
effect~\cite{bricard_emergence_2013,bricard_emergent_2015} as well as macroscopic collective motion patterns in
 bird flocks or sheep herds~\cite{vicsek_collective_2012,ginelli_intermittent_2015,toulet_imitation_2015}. 
In particular, the study of bacterial systems as well as their theoretical analysis within simple 
self-propelled particle models has lead to interesting insights into the physics of active 
matter~--~consider, for example, the clustering 
of myxobacteria~\cite{peruani_collective_2012,peruani_nonequilibrium} or the dynamic vortex formation in 
dense suspensions of swimming bacteria~\cite{wensink_meso_2012,dunkel_fluid_2013,grossmann_vortex_2014,grossmann_pattern_2015}.

In order to understand the cooperative behavior of active particles as well as the associated pattern 
formation processes, reliable knowledge of the dynamics of individual entities is crucial. 
In this work, we therefore focus on the dynamics of individual active particles.
We particularly consider particles that are able to reverse their direction of motion repeatedly. 
More precisely, particles follow an alternating motion pattern where rather persistent motion is interrupted 
by sudden reversals of the direction of motion.

This type of motion has been reported in a variety of bacterial systems~\cite{Wu_periodic_2009,thutupalli_directional_2015,johansen_variability_2002,Barbara_bacterial_2003,beer_periodic_2013,duffy_turn_1997,davis_2d_2011,theves_bacterial_2013,raatz_swimming_2015}.  
For instance, the soil bacterium \textit{Myxococcus xanthus} constitutes a paradigmatic example of a bacterium exhibiting periodic reversals
in the direction of motion:~internal oscillations of the protein dynamics cause switches in cell polarity and, correspondingly, in 
the direction of motion~\cite{leonardy_reversing_2008,Wu_periodic_2009,rashkov_model_2012,thutupalli_directional_2015}. 
%
%
Under certain conditions, the reversals of several, densely packed bacteria appear synchronously leading to 
remarkable accordion wave patterns~\cite{boerner_rippling_2002,sliusarenko_accordion_2006}. 
Apart from myxobacteria, a variety of marine microorganisms exhibit \textit{run-\&-reverse}
motion~\cite{johansen_variability_2002}, such as
\textit{Pseudoalteromonas haloplanktis} and \textit{Shewanella putrefaciens}~\cite{Barbara_bacterial_2003}. 
Similar motion patterns were reported for \textit{Pseudomonas citronellolis}~\cite{taylor_reversal_1974}, \textit{Paenibacillus
dendritiformis}~\cite{beer_periodic_2013} and \textit{Pseudomonas
putida}~\cite{duffy_turn_1997,davis_2d_2011,theves_bacterial_2013,raatz_swimming_2015}. 

More complex, three-step (\textit{run-reverse-flick}) motion patterns,  composed of rather straight runs, directional 
reversals and $90^\circ$ turns, were found in the marine bacterium 
\textit{Vibrio alginolyticus}~\cite{xie_bacterial_2011,Stocker_reverse_2011}.
In this context it has been speculated that bacteria can adopt their flip and reversal frequencies to the 
environmental conditions thus affecting their chemotactic response in order to detect and climb up chemical 
gradients more efficiently~\cite{xie_bacterial_2011}. 
Similar questions were addressed theoretically within the context of self-propelled particle models -- the chemotactic drift of
self-propelled particles with run-reverse-flick motility was studied
in~\cite{taktikos_motility_2013}. 

Recently, the diffusion properties of a class of active particles performing \textit{run-\&-turn} motion -- a motion pattern where
persistent continuous runs are interrupted by sudden reorientation events occurring after stochastic waiting times -- were generally
derived in \cite{detcheverry_nonpoissonian_2015} by means of noncommuting operators. Active particle with reversal are a special case of
this \textit{run-\&-turn} motility pattern. 
Further, the influence of speed fluctuations on the diffusion of active particles with directional reversal in one spatial dimension were
investigated in the context of \textit{active Brownian particles}~\cite{romanczuk_active_2011,romanczuk_active_2012}

In this work, we study the diffusion properties of self-driven particles, particularly focusing on the effect of directional 
reversals. 
Generally, the microscopic dynamics of active particles results from the complex interplay of multiple 
factors:~particle shape, detailed properties of the propulsion mechanism, interaction with the surroundings, 
e.g.~hydrodynamic interaction with a fluid or friction on a surface, etc. 
Here, we abstain from modeling the details of the self-propulsion mechanism and reduce the complexity of the  
biochemical processes involved in the reversal events to a simple clock model as discussed in detail below.  
In short, we consider a minimalistic self-propelled particle model which includes the following basic mechanisms into the 
dynamics: 
\begin{enumerate}
 \item[(i)] the propulsive force enabling a particle to move actively is counterbalanced by friction leading
to active motion at a non-vanishing, constant, characteristic speed~$v_0$;
 \item[(ii)] the trajectories of particles are not perfectly straight lines due to fluctuations of the direction of the driving force
or spatial heterogeneities, which is taken into account by the addition of rotational noise;  
 \item[(iii)] reversal events:~recurrent switching of the internal motor between two states that correspond to \textit{forward} 
and \textit{backward} motion. 
\end{enumerate}
We exclusively consider homogeneous spatial environments without addressing questions related to chemotaxis. 

This work is structured as follows. 
Section~\ref{sec:APWR} introduces a paradigmatic model for the spatial dynamics of active particles with directional 
reversal. 
Moreover, central quantities of interest are defined and their interrelation is discussed. 
In particular, we derive a general expression for the diffusion coefficient of active particles with 
reversal.  
In section~\ref{sec:ClockModel}, we introduce a stochastic clock model representing the intracellular 
biochemical cascade triggering reversal events. 
This simple model allows us to reproduce the characteristic shape of the distribution of times in between two 
consecutive reversal events as observed in experiments.
We use this clock model to illustrate characteristic properties of the velocity correlation 
function, mean squared displacement and the diffusion coefficient of active particles with reversal. 
In section~\ref{sec:renewal}, we extend the clock model describing the reversal dynamics by a renewal process. 
The analysis of this general model for reversing self-propelled particles reveals that~--~under certain
circumstances~--~the diffusion coefficient exhibits a maximum at a finite rotational noise intensity.
This resonance effect is explained in detail and the relevance of this finding for bacterial systems is addressed by comparing theoretical
predictions with experimental measurements. 
An outlook~--~accompanied by a summary of our main results~--~is given in the last section.

\section{Active particles with reversal}
\label{sec:APWR}

\begin{figure}[t]
   \begin{center}
     \includegraphics[width=0.685\columnwidth]{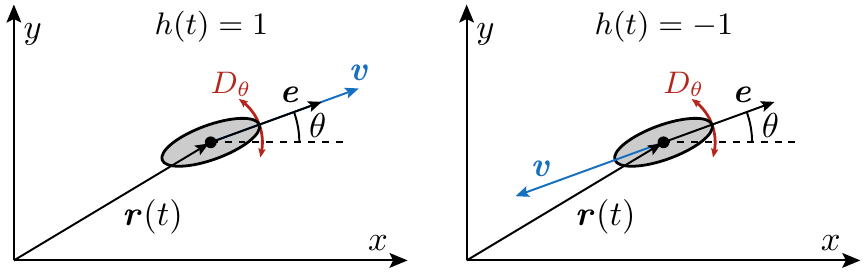}
     \caption{Schematic visualization of a particle at position $\vec{r}(t)$. Its body axis is
oriented along the unit vector $\vec{e}$ (black arrow). \textit{Left}:~The motility engine is in state $h(t) = 
1$;~\textit{right}:~the propulsion engine is in the opposite state, $h(t) = -1$. The velocity of a particle, 
depicted by a blue arrow, is determined by $\vec{v}(t) = v_0 \vec{e}(t) h(t)$. The red arrows indicate 
stochastic rotation of the body axis due to external inhomogeneities or fluctuations of the propelling engine  
modeled by white noise with intensity $D_\theta$.  }
     \label{fig:ParticleScheme}
    \end{center}
  \end{figure}

\subsection{Dynamics in general}

In the following, we describe the mathematical model:~self-propelled particles with directional reversal.
Active particles, e.g.~bacteria, exhibit a body axis, which we denote by an unit vector 
$\vec{e}(t)$~(see also
Fig.~\ref{fig:ParticleScheme}). 
Suppose that  the propulsion engine of a particle switches between two states in a cyclic manner implying 
alternating parallel (forward) and antiparallel (backward) motion with respect to this axis. 
The two states of the internal motor are reflected by a state function $h(t) \in \{ -1 , 1\}$.  
A reversal event is then described by the transition 
 \begin{eqnarray}
  \label{eqn:mod:reversal} 
   h(t) \; \rightarrow \; - h(t). 
 \end{eqnarray}
This process is assumed to be fast compared to the mean time in between two reversals as observed
experimentally~\cite{theves_bacterial_2013,Wu_periodic_2009}. The time between transitions is modeled by a clock model, which is defined 
in the next section.     

The velocity $\vec{v}(t)$ of a particle is determined by the product of the characteristic speed $v_0$~--~we 
assume a stationary force balance of driving and drag forces neglecting speed fluctuations~--, the unit vector 
$\vec{e}(t)$ indicating
the orientation of the body axis as well as the state of the propelling engine $h(t)$. 
Therefore, the spatial dynamics of a self-propelled particle with directional reversal reads 
 \begin{eqnarray}
  \label{eqn:model:space}
  \frac{\rmd \vec{r}(t)}{\rmd t} &= \vec{v}(t) = v_0 \vec{e}(t) h(t) ,
 \end{eqnarray} 
where $\vec{r}(t)$ hereafter denotes the position in space\footnote{This model equivalently describes 
particles which perform abrupt $180^\circ$ turns instead of
reversals. The function $h(t)$ is a bookkeeping parameter serving as a convenient description of turns in this 
case. }.
We assume that the orientation of the body axis fluctuates stochastically due to spatial heterogeneities or noise associated to
the self-propelling engine. 
This random reorientation is taken into account by addition of rotational diffusion. 
Here, we focus on the motion in two spatial dimensions, i.e.~on substrates~--~the most relevant experimental 
setup. 
In two dimensions, the orientation of the body axis is determined by a time-dependent polar angle
$\theta(t)$, cf.~Fig.~\ref{fig:ParticleScheme} for an illustration. 
The temporal dynamics of the body axis, parametrized by the polar angle, reads
 \begin{eqnarray}    
   \vec{e}(t) = \left(\begin{array}{@{}c@{}}  \cos \theta(t) \\ \sin \theta(t) \end{array}\right), \quad    
   \frac{\rmd \theta(t)}{\rmd t} = \sqrt{2 D_\theta} \, \xi(t).  
 \end{eqnarray}
The random process $\xi(t)$ denotes Gaussian white fluctuations with zero mean, $\mean{\xi(t)} = 0$, and 
temporal $\delta$-correlations:~$\mean{\xi(t) \xi(t')} = \delta(t - t')$.

The noise intensity $D_\theta$ is inversely proportional to the persistence length $l_p \sim v_0/D_\theta$ of 
the trajectory of an active particle.
In general, $D_\theta$ itself may depend on additional parameters such as the 
speed itself~\cite{romanzuk_brownian_2011}.
However, we will treat it as an independent parameter in this context.

As discussed below, our results are not restricted to two dimensional systems since the motion of a 
self-propelled particle in two dimensions is not fundamentally different from corresponding three (or higher)
dimensional cases~\cite{grossmann_anistropic_2015}. 
We will comment in the respective paragraphs below which findings do quantitatively change in dimensions 
larger than two. 
Note that our analysis automatically contains the one-dimensional motion of self-propelled particles with
directional reversal:~the back and forth motion along a line is recovered in the zero noise
limit~($D_\theta~=~0$) in our model.  

\subsection{Velocity correlation, mean squared displacement and diffusion coefficient}
\label{sec:gen:dif:prop}

In this section, we define several important observables used to characterize the motion pattern of 
self-propelled particles and briefly discuss their interrelation. 
Further, an expression for the diffusion coefficient of self-propelled particles with reversal is 
derived.

The central quantity of interest is the correlation function of the velocity:
 \begin{eqnarray}
  \label{eqn:vel:corr:func}
   \mean{\vec{v}(t) \cdot \vec{v}(t')} = v_0^2 \mean{\vec{e}(t) \cdot \vec{e}(t')} \mean{h(t)h(t')}. 
 \end{eqnarray}
In the expression above, we assumed stochastic independence of the temporal dynamics of the body axis 
$\vec{e}(t)$ and the occurrence of reversal events. 
This is a reasonable assumption since both processes are of different physical origin.
Due to the stochastic independence, the calculation of the velocity correlation function can be done in two 
subsequent steps, considering the dynamics of the body axis and the reversal dynamics separately.

We point out that the correlation function of the body axis, $\mean{\vec{e}(t) \cdot \vec{e}(t')}$, is 
equal to the corresponding correlation function of a self-propelled particle without reversal.
This limit is recovered from Eq.~\eqref{eqn:vel:corr:func} by setting $h(t) = 1$ for all times.  
This correlation function is known to decay exponentially as discussed 
in~\cite{schienbein_langevin_1993,mikhailov_self_1997}: 
 \begin{eqnarray}
   \mean{\vec{e}(t) \cdot \vec{e}(t')} = e^{-D_\theta \abs{t - t'}} .
 \end{eqnarray}
Thus, the velocity correlation function reads 
 \begin{eqnarray}
   \label{eqn:velo:corr:func:7}
   \mean{\vec{v}(t) \cdot \vec{v}(t+\tau)} = v_0^2 \, e^{-D_\theta \tau} C_{hh}(t,\tau), 
 \end{eqnarray}
where $\tau>0$ and $C_{hh}(t,\tau) = \mean{h(t) h(t+\tau)}$. 
Due to the exponentially decaying envelope, the velocity correlation function does not possess heavy tails for 
any nonzero noise~$D_\theta$ excluding superdiffusion a priori irrespective of $C_{hh}(t,\tau)$. 
Subdiffusion is neither expected for finite noise amplitudes. 
Therefore, the mean squared displacement and, in particular, the diffusion coefficient are sufficient to 
characterize the long-term motion. 
Qualitatively, these arguments hold in higher spatial dimensions as well:~In $d$ dimensions, the 
correlation functions decays according to $\mean{\vec{e}(t) \cdot \vec{e}(t')} =
e^{-D_\varphi (d-1) \abs{t - t'}}$. Thus, the correlation time is affected by the spatial dimensionality
only. However, the qualitative exponential decay exists in all dimensions~\cite{grossmann_anistropic_2015}. 

The mean squared displacement is directly related to the velocity correlation function via the following 
double integral, known as Taylor-Kubo formula~\cite{Taylor_diffusion_1922,kubo_statistical_1957} 
  \begin{eqnarray} 
  \label{eqn:Tailor:Kubo:relation}
    \mean{\abs{\vec{r}(t) - \vec{r}(0)}^2} 
    &= 2 \int_{0}^t \! \rmd t' \int_0^{t'} \! \rmd t'' \, \mean{\vec{v}(t') \cdot \vec{v}(t'')} \! ,
   \end{eqnarray}
which is proved by direct integration of Eq.~\eqref{eqn:model:space} and using the symmetry of the velocity
correlation function with respect to permutation of the times $t'$ and $t''$. 
The asymptotic spatial diffusion coefficient $\mathcal{D}$ follows, in turn, from the mean squared 
displacement via 
 \begin{eqnarray}
  \label{eqn:def:diff:coeff}
  \mathcal{D} &= \frac{1}{4} \cdot \lim_{t\rightarrow \infty} \left [ \frac{\rmd \! \mean{\abs{
  \vec{r}(t) - \vec{r}(0)}^2}}{\rmd t} \right ] \! .
 \end{eqnarray}
This definition can be rewritten by making use of the Taylor-Kubo relation. 
Subsequent insertion of the velocity correlation function, Eq.~\eqref{eqn:velo:corr:func:7}, finally yields 
 \begin{eqnarray}
  \mathcal{D} &= \frac{v_0^2}{2} \cdot \int_0^{\infty} \! \rmd \tau \, e^{-D_\theta \tau}
\lim_{t\rightarrow \infty} \left [C_{hh}(t-\tau,\tau) \textcolor{white}{\frac{a}{b}} \! \! \! \!\! \right] \! 
.
 \end{eqnarray}
We denote the correlation function in the limit $t\rightarrow\infty$ by
 \begin{eqnarray}
  C_{hh}^{(eq)}(\tau) = \lim_{t\rightarrow \infty} \left [C_{hh}(t-\tau,\tau) \textcolor{white}{\frac{a}{b}} 
\! \! \! \!\!
\right] \! .
 \end{eqnarray}
Consequently, the diffusion coefficient is determined by the integral transform
  \begin{eqnarray}
    \mathcal{D} &= \frac{v_0^2}{2} \int_0^{\infty} \! \rmd \tau \, e^{-D_\theta \tau} \, C^{(eq)}_{hh} \! 
\left 
(\tau \right ) 
		= \frac{v_0^2}{2} \, \widehat{C}^{(eq)}_{hh} \! \left (D_\theta \right ) \!. 
\label{eqn1:diff:coeff}
  \end{eqnarray}   
Interestingly, the integral in Eq.~\eqref{eqn1:diff:coeff} is structurally equivalent to the Laplace 
transform\footnote{We use the definition $\widehat{f}(s)=\int_0^{\infty} \rmd t \, e^{-st} f(t)$ for the 
Laplace 
transform $\widehat{f}(s)$ of a function $f(t)$~\cite{doetsch_tabellen_1947}.} of the correlation function 
$C^{(eq)}_{hh} \! \left (\tau \right )$.
This central result is convenient for the calculation of the diffusion coefficient because the 
correlation function of the process~$h(t)$ can be calculated in the Laplace domain whereas the inverse 
transformation is often impossible. 

We conclude this section by giving the diffusion coefficient for self-propelled particles in spatial 
dimensions $d\ge 2$ which is similar in structure to the two dimensional result:  
 \begin{eqnarray}
  \label{eqn2:diff:coeff}
  \mathcal{D} = \frac{v_0^2}{d} \, \widehat{C}^{(eq)}_{hh} \! \left ( D_\theta (d-1) 
\textcolor{white}{\frac{a}{b}} \! \! \! \!
\right ) \!. 
 \end{eqnarray}
Thus we obtain the concise result that the diffusion coefficient of a self-propelled particle with directional 
reversal is determined by the Laplace transform of the correlation function of the reversal process $h(t)$. 

\section{A clock model}
\label{sec:ClockModel}

\subsection{Directional reversal controlled by an intracellular clock}
\label{subsec:IntoCM}

The triggering of reversal events is controlled by complex processes taking place inside a particle giving 
rise to stochastic occurrences of reversals. 
Here, we do not model the internal particle dynamics in detail since these processes are hardly accessible 
experimentally anyway.
We rather employ a coarse-grained description capturing the essential phenomenology of the reversal dynamics.

\begin{figure}[b]
  \begin{center}
  \includegraphics[width=0.75\columnwidth]{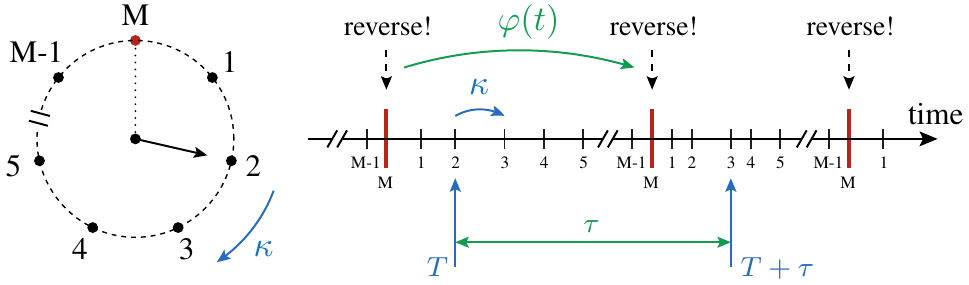}   
  \end{center}
  \caption{Schematic visualization of the internal particle dynamics modeled by a stochastic clock with $M$ 
ticks. The black arrow indicates the clock hand which jumps in a small time interval $\Delta t$ with 
probability $\kappa \Delta t$. Thus, the ticking of the clock is a Poisson process with rate $\kappa$. 
Whenever the clock hand crosses the state $M$, marked in red, a reversal process is triggered. 
The timeline on the right symbolizes the same process as a function of time. The time intervals in between 
two reversal events are distributed according to the probability density $\varphi(t)$, 
Eq.~\eqref{eqn:WTPDF:Clock}. Furthermore, we indicate the state of the clock at a time $T$ (state $2$). 
After a time $\tau$, a reversal may have occurred~(in Fig.:~one reversal) and the clock is in another 
internal state~(in Fig.:~state~$3$). }
  \label{fig:ClockScheme_WTPDF}
 \end{figure}

In order to trigger a single reversal, a certain number of biochemical (activation) processes needs to be 
executed.  
Following this reasoning, we propose a \textit{stochastic clock model} (cf.~Fig.~\ref{fig:ClockScheme_WTPDF}) 
which is intended to represent the internal particle dynamics. 
Suppose that each of the activation processes~--~corresponding to ticks of the clock~--~arises at a given rate 
which we assume to be all identical for simplicity. 
Whenever the watch hand completes a full revolution, i.e.~$M$ consecutive ticks appeared, a reversal event is 
triggered. 

We model the ticking of the clock as a stochastic process:~the watch hand ticks with a probability $\kappa 
\Delta t$ in a small time increment $\Delta t$. 
Thus, the ticking of the clock is described, mathematically speaking, by a Poisson process~\cite{gardiner_stochastic_2010} with
rate~$\kappa$. 
This stochastic process is unique since, as has already been mentioned, the probability that a tick is 
observed in a given time interval is constant and, hence, independent of the process history. 

\begin{figure}[t]
  \begin{center}
  \includegraphics[width=0.45\columnwidth]{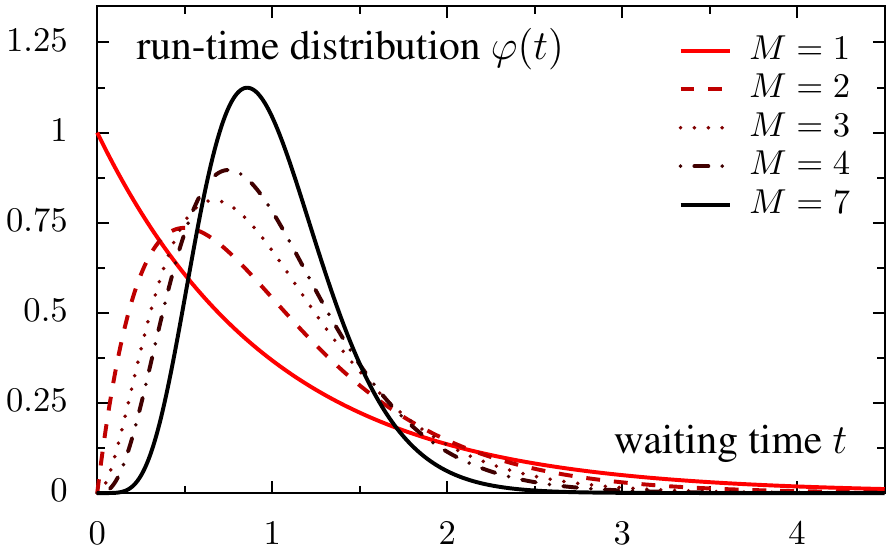}   
  \end{center}
  \caption{Run-time distribution~\eqref{eqn:WTPDF:Clock} resulting from the clock model for several values of 
$M$ (\#$\,$ticks of the clock). The run-time distribution reduces to an exponential distribution for $M=1$. In 
contrast, $\varphi(t)$ tends towards a Gaussian bell-shape for large values of~$M$. The rate $\kappa = M$ was 
adjusted such that the mean of $\varphi(t)$ equals one:~$\mean{t} = \int_0^\infty \rmd t \, t \,\varphi(t) 
\overset{!}{=} 1$. }
  \label{fig:WTPDF:Clock}
 \end{figure}

Whereas the biochemical processes controlling the reversals are not directly observable, the resulting 
distribution of the times elapsed in between two successive reversal events~--~usually called \textit{run-time 
distribution}~--~is easily accessible experimentally. 
We will denote the run-time distribution by~$\varphi(t)$. 
The clock model introduced above implies one particular run-time distribution ($\gamma$-distribution) which
reads  
 \begin{eqnarray} 
  \label{eqn:WTPDF:Clock}
  \varphi(t) = \frac{\kappa^M t^{M-1} e^{-\kappa t} }{\left (M-1 \right )!} . 
 \end{eqnarray}
Naturally, $\varphi(t)$ depends on two parameters:~the number of ticks~$M$ of the clock as well as the 
rate~$\kappa$ at which ticks of the watch hand are observed. 
The distribution is plotted for several values of $M$ in Fig.~\ref{fig:WTPDF:Clock}. 
The limiting case $M=1$ is special since the clock possesses only one tick, such that every tick of the clock 
implies a reversal event. 
Accordingly, the occurrence of reversals is a Poisson process and $\varphi(t)$ is an exponential distribution 
in this case. 
In contrast, an asymmetric bell shape is observed for $M>1$.
In the limit of large $M$, $\varphi(t)$ tends towards a Gaussian distribution. 
We comment on the applicability of the run-time distribution $\varphi(t)$ in section~\ref{sec:summary} by a 
direct comparison to experimentally observed distributions. 

The most important characteristics of the run-time distribution are its mean $\mean{t}$ determining the 
average frequency $\lambda_r = 1/\!\mean{t}$ at which reversal events are observed as well as its width 
$\sigma^2 = \mean{\left ( t - \mean{t} \right )^2}$. 
The mean time separating two reversal events is equal to the $M$-fold of the mean waiting time for a single 
tick on the clock determined by $\kappa^{-1}$. 
Therefore, the reversal frequency is given by $\lambda_r = \kappa / M$. 
The variance $\sigma^2$ of the gamma distribution, Eq.~\eqref{eqn:WTPDF:Clock}, reads $\sigma^2 = M/\kappa^2$.
Hence, the coefficient of variation $c_v$, i.e.~the standard deviation over the mean, decreases with the 
number of intermediate
steps:~$c_v = 1/\sqrt{M}$. 
Consequently, the interpretation of the parameters of the clock model is straightforward:~the accuracy of the 
clock,
i.e.~the regularity at which reversal events occur~--~reflected by the width of the run-time distribution 
$\varphi(t)$~--~is
determined by the number of ticks $M$, whereas the ticking rate $\kappa$ is directly proportional to the mean 
reversal frequency $\lambda_r$. 

\subsection{Analysis of the clock model}
\label{subsec:anaCM}

Starting from the clock model, the calculation of the correlation function $C_{hh}(t,\tau)$ is sketched.  
Subsequently, the resulting phenomenology of this clock model is discussed. 

The reversal process $h(t)$ which determines the state of the propelling engine of a particle does only take
two values corresponding to \textit{forward} and \textit{backward} motion:~$h(t) \in \{ -1 ,1 \}$. 
Therefore, the product $h(t) \cdot h(t+\tau)$ is equal to plus or minus one depending on the number of 
reversal events in the
time window $\tau$ beginning at time~$t$. 
Accordingly, we can calculate the correlation function $C_{hh} (t,\tau) = \mean{h(t) h(t+ \tau)}$ of the 
process $h(t)$ via
\begin{eqnarray}
 \label{eqn2:corr:func:def2}
 C_{hh} (t,\tau) = \mean{h(t) h(t+\tau)} = \sum_{N=0}^{\infty} (-1)^N P_N(t,\tau), 
\end{eqnarray}
where $P_N(t,\tau)$ denotes the probability that exactly $N$ reversals occurred within the time 
interval~$\tau$. 

The clock model yields an illustrative way to calculate the probabilities $P_N(t,\tau)$, 
cf.~Fig.~\ref{fig:ClockScheme_WTPDF} for a visualization. 
We make use of the fact that the ticking of the clock is equivalent to a Poisson process with rate $\kappa$. 
For a Poisson process, the probability $\psi_n(t)$ to observe $n$ ticks of the clock in a given time interval 
of length~$t$ is determined by 
\begin{eqnarray}
  \frac{\rmd \psi_n(t)}{\rmd t} = \left\{\begin{array}{@{}l@{\quad}l}
      - \kappa \psi_{0}(t), & n=0, \\[\jot]
      - \kappa \psi_{n}(t) + \kappa \psi_{n-1}(t) , & n \ge 1,
    \end{array}\right. 
\end{eqnarray}
whose solution is found by successive integration: 
\begin{eqnarray}
 \psi_n(t) = \frac{(\kappa t)^n e^{-\kappa t}}{n!} . 
\end{eqnarray}
Using this result, the probability to observe no reversal within a time interval of length~$\tau$ beginning 
at $t$ can be written as 
\begin{eqnarray}
 \label{eqn:prob.0rev:c}
 P_0(t,\tau) &= \underbrace{\sum_{N=0}^{\infty} \sum_{n=0}^{M-1} \, \psi_{NM+n}(t) \! \!}_{(a)} 
\underbrace{\sum_{m=0}^{M-n-1}
\! \psi_m(\tau)}_{(b)} . 
 \end{eqnarray}
This equation consist of two parts to be interpreted as follows.  
The terms $(a)$ reflect the state of the clock at time~$t$.
The clock is in one of the $M$ internal states, denoted by~$n$.
Previous to time $t$, a certain number of reversals~$N$ have already occurred. 
Thus, $NM+n$ ticks of the clock were observed up to time $t$ in total. However, the number of reversals and 
the state of the
internal clock are not relevant~--~that is why $N$ and $n$ is to be summed over. 
The third sum, abbreviated by $(b)$ in Eq.~\eqref{eqn:prob.0rev:c}, reflects the probability that $m$ 
additional ticks of the
clock are observed. 
The set of $m$ values is constrained to those values satisfying $n+m<M$, i.e.~such that no reversal event 
occurs within the time
interval $\tau$. 
Pictorially speaking, this is fulfilled if the watch hand does not complete a revolution within~$\tau$~(see 
left of
Fig.~\ref{fig:ClockScheme_WTPDF}). 

Along similar lines of arguments, the probability to observe $k$ reversals within time~$\tau$ can be derived: 
\begin{eqnarray}
 \label{eqn:prob.krev:c}
 P_k(t,\tau) &= \sum_{N=0}^{\infty} \sum_{n=0}^{M-1} \psi_{NM+n}(t) \! \! \sum_{m=0}^{M-1} 
\psi_{Mk+m-n}(\tau). 
\end{eqnarray}
Apparently, the first part does not change whereas the third sum is replaced by the probabilities that exactly 
$k$ reversals
occur within $\tau$ corresponding to $k$ revolutions of the watch hand. 

Eqs.~\eqref{eqn2:corr:func:def2}--\eqref{eqn:prob.krev:c} allow the calculation of the correlation function 
by performing the summations.
In time domain, this is not possible in general. 
However, the summation can be done in a closed form in Laplace domain by inserting the Laplace transform 
of~$\psi_n(t)$, 
\begin{eqnarray}
 \widehat{\psi}_n(s) = \frac{\kappa^n}{(\kappa + s)^{n+1}},
\end{eqnarray}
and summing several geometrical series. 
The probabilities $P_0$ and $P_k$ read in Laplace space as follows:
\eqnlabel{eqn:probab:jumps}
\numparts
\begin{eqnarray}
 \!\!\!\widehat{\widehat{P}\,}_{\!\!0}(s,u) &= \frac{1}{u} \! \cdot \! \left [ \frac{1}{s} - \frac{1}{1 - 
\widehat{\varphi}(s)}
 \! \cdot \! \frac{\widehat{\varphi}(s) -
\widehat{\varphi}(u)}{u - s} \right
] \! ,\\
 \!\!\!\widehat{\widehat{P}\,}_{\!\!k}(s,u) &= \frac{1 - \widehat{\varphi}(u)}{u} \! \cdot \! \frac{\left [ 
\widehat{\varphi}(u)
\right ]^{k-1}}{1 - \widehat{\varphi}(s)} \! \cdot \! \frac{\widehat{\varphi}(s) - \widehat{\varphi}(u)}{u - 
s} .
\end{eqnarray}
\endnumparts
Finally, we also obtain a closed expression for the correlation function: 
 \begin{eqnarray}
   \label{eqn:gen:corr:func:laplace2}
   \widehat{\widehat{C}\,}_{\!\!hh}(s,u) = \frac{1}{u} \!\cdot\! \left [ \frac{1}{s} - \frac{2}{u-s} \!\cdot\!
\frac{\widehat{\varphi}(s) -
\widehat{\varphi}(u)}{ \left [1 + \widehat{\varphi}(u) \right ] \! \cdot \! \left [1 - \widehat{\varphi}(s) 
\right ] } \,
\right ] \! . \! \!
 \end{eqnarray}
In Eq.~\eqref{eqn:probab:jumps} and~\eqref{eqn:gen:corr:func:laplace2}, the Laplace transform 
\begin{eqnarray}
 \widehat{\varphi}(s) = \left [ \frac{\kappa}{\kappa + s} \right ]^M
\end{eqnarray}
of the run-time distribution $\varphi(t)$, cf.~Eq.~\eqref{eqn:WTPDF:Clock}, was identified and abbreviated for 
convenience.  
In general, the correlation function depends on both, $t$ and $\tau$, via $s$ and $u$ in Laplace domain~--~the
reversal dynamics $h(t)$ possesses memory reflecting the non-Markovian~\cite{gardiner_stochastic_2010} 
character of the dynamics. 
For the long-time diffusion properties, however, only the limiting behavior $\lim_{t\rightarrow \infty} \left 
[C_{hh}(t,\tau)
\textcolor{white}{\frac{a}{b}} \! \! \! \! \right] = C_{hh}^{(eq)}(\tau)$ is relevant. 
The Laplace transform of $C_{hh}^{(eq)}(\tau)$ is found from the general solution, 
Eq.~\eqref{eqn:gen:corr:func:laplace2}, as
follows~\cite{doetsch_tabellen_1947}:
\eqnlabel{eqn:red:corr:func:4}
\numparts
\begin{eqnarray}
   \widehat{C}_{hh}^{(eq)}(u) &= \lim_{t \rightarrow \infty} \left [ \widehat{C}_{hh}(t,u) \right ] = \lim_{s 
\rightarrow 0} \left
[ s \,
\widehat{\widehat{C}\,}_{\!\!hh}(s,u) \right ] \\
  &= \frac{1}{u} \! \cdot \! \left [ 1 - \frac{2 \lambda_r}{u} \! \cdot \! 
\frac{1 - \widehat{\varphi}(u)}{1 +
\widehat{\varphi}(u)} \, \right ] \! . 
 \end{eqnarray}
\endnumparts
Consequently, the central quantity of interest, the correlation function 
$C^{(eq)}_{hh}(\tau)$ determining the diffusion properties of an active particle whose reversal is 
triggered by an internal clock, is known. 

\subsection{Results: illustration of the clock model}
\label{sec:subsec:res}

In the following, we illustrate our results and outline their implications.
We begin the discussion of the motion characteristics by recalling the general form of the velocity 
correlation function: 
 \begin{eqnarray}
  \label{eqn:velo:corr:func:100}
   \mean{\vec{v}(t) \cdot \vec{v}(t+\tau)} = v_0^2 \, e^{-D_\theta \tau} C_{hh}(t,\tau).  
 \end{eqnarray}
For the self-propelled particle model considered here, the velocity correlation increases proportional to the 
squared speed. 
The rotational noise inducing a stochastic rotation of the body axis implies an exponential damping of the 
velocity correlations. 
The characteristic damping time $\tau_\theta$ is inversely proportional to the amplitude of rotational 
fluctuations:~$\tau_\theta = 1/D_\theta$. 
Accordingly, reversal events do only play an important role if the mean time $\mean{t}$ 
separating subsequent reversal events is smaller than or comparable to the correlation time~$\tau_\theta$. 
In the opposite case, 
rotational noise causes the body axis to rotate within a shorter time compared to $\mean{t}$ implying the 
decorrelation of
velocities within this time. 
To be concrete, let us suppose that reversal events occur at times $t_1$ and~$t_2$. If the velocities 
$\vec{v}(t_1)$
and $\vec{v}(t_2 - \epsilon)$, i.e.~immediately before the next reversal event ($\epsilon \ll 1$), had already 
been
uncorrelated, the reversal at time $t_2$ does not influence the velocity statistics. 
Consequently, the diffusion properties are independent of the directional reversals in this regime. 

\begin{table}[b]
\caption{Correlation functions $C_{hh}^{(eq)}(\tau)$ for the clock model.}
\label{tab:corrFunc}
\begin{indented}
\item[]\begin{tabular}{@{}cl}
\br
number of ticks & correlation function \\
\mr
$M=1$: & $C_{hh}^{(eq)}(\tau) = e^{-2 \kappa \tau}$ \\
$M=2$: & $C_{hh}^{(eq)}(\tau) = e^{-\kappa \tau} \cos (\kappa \tau)$  \\
$M=3$: & $C_{hh}^{(eq)}(\tau) = \frac{1}{9} \! \left [ e^{-2 \kappa \tau} + 8 \!\,e^{-\frac{\kappa \tau}{2}} 
\cos
\! \left ( \! \frac{\sqrt{3} \kappa \tau}{2} \right ) \! \right ]$ \\
$M=4$: & $C_{hh}^{(eq)}(\tau) = e^{-\kappa \tau} \cos \! \left ( \! \frac{\kappa \tau}{\sqrt{2}} \right ) \! 
\left [ \cosh \!
\left ( \! \frac{\kappa \tau}{\sqrt{2}} \right ) \!+\! \frac{1}{\sqrt{2}} \sinh \! \left ( \! \frac{\kappa 
\tau}{\sqrt{2}} \right
) \right ]$ \\
\br
\end{tabular}
\end{indented}
\end{table}

 \begin{figure}[t]
  \begin{center}
\includegraphics[width=0.45\columnwidth]{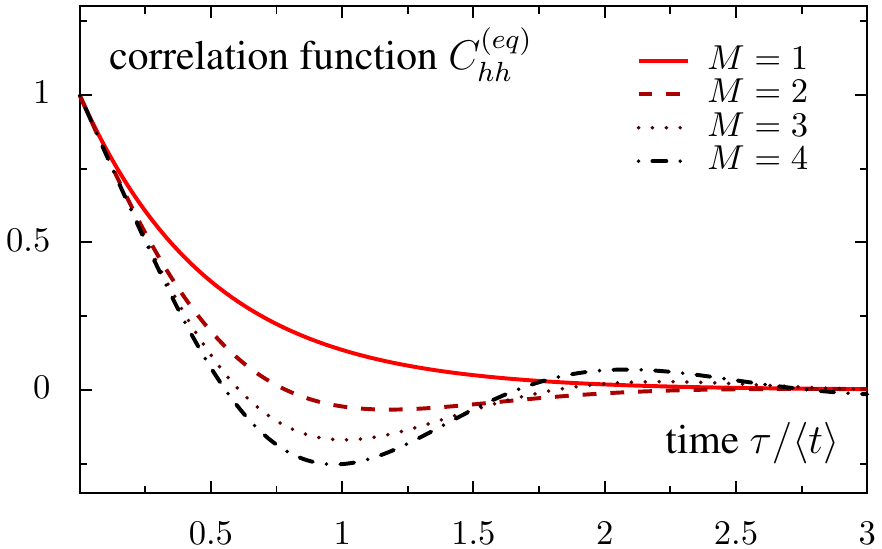}
  \end{center}
  \caption{Correlation function $C_{hh}^{(eq)}(\tau)$, cf.~Tab.~\ref{tab:corrFunc}, as a function of time 
$\tau$ for the clock
model. The number of ticks of the clock is denoted by $M$. For $M\ge2$, oscillations of the correlation 
function are observed.
The limit $M=1$ is special since the reversal process reduces to a Poisson process in this case. That is why 
correlations decay
exponentially for $M=1$. }
  \label{fig:Corr_Func}
 \end{figure}

The velocity correlation function is proportional to the correlation function of the reversal process $h(t)$ 
which was
calculated in the preceding section for the clock model. 
The correlation function crucially depends on the number of ticks $M$ of the clock. 
In Tab.~\ref{tab:corrFunc}, we summarize the correlation function $C_{hh}^{(eq)}(\tau)$ for the lowest $M$ 
values which are
also graphically shown in Fig.~\ref{fig:Corr_Func}.  
For $M=1$, the reversal process reduces to a Poisson process since every tick of the clock triggers a reversal 
event which
therefore occur at a constant rate $\kappa$. 
Accordingly, the correlation function is exponentially decaying with a characteristic time determined by 
$1/\kappa$. 
More interesting behavior is observed if the clock possesses several ticks, $M\ge2$:~the correlation function 
does oscillate. 
Oscillations become more and more pronounced with increasing $M$. 
Remember that the number of ticks of the clock controls the accuracy of reversal event occurrences.
In the limit $M \rightarrow \infty$, reversals would occur deterministically every $\mean{t} = 1/\lambda_r$ implying a square wave signal
form of the correlation function~$C_{hh}^{(eq)}(\tau)$. 

Oscillations are likewise expected for the velocity correlation function, Eq.~\eqref{eqn:velo:corr:func:100}, 
in the low noise
regime~($D_\theta~\lesssim~\lambda_r$) where oscillations reflect the recurrent back and forth motion of 
particles.

 \begin{figure}[b]
  \begin{center}
   \includegraphics[width=0.8\columnwidth]{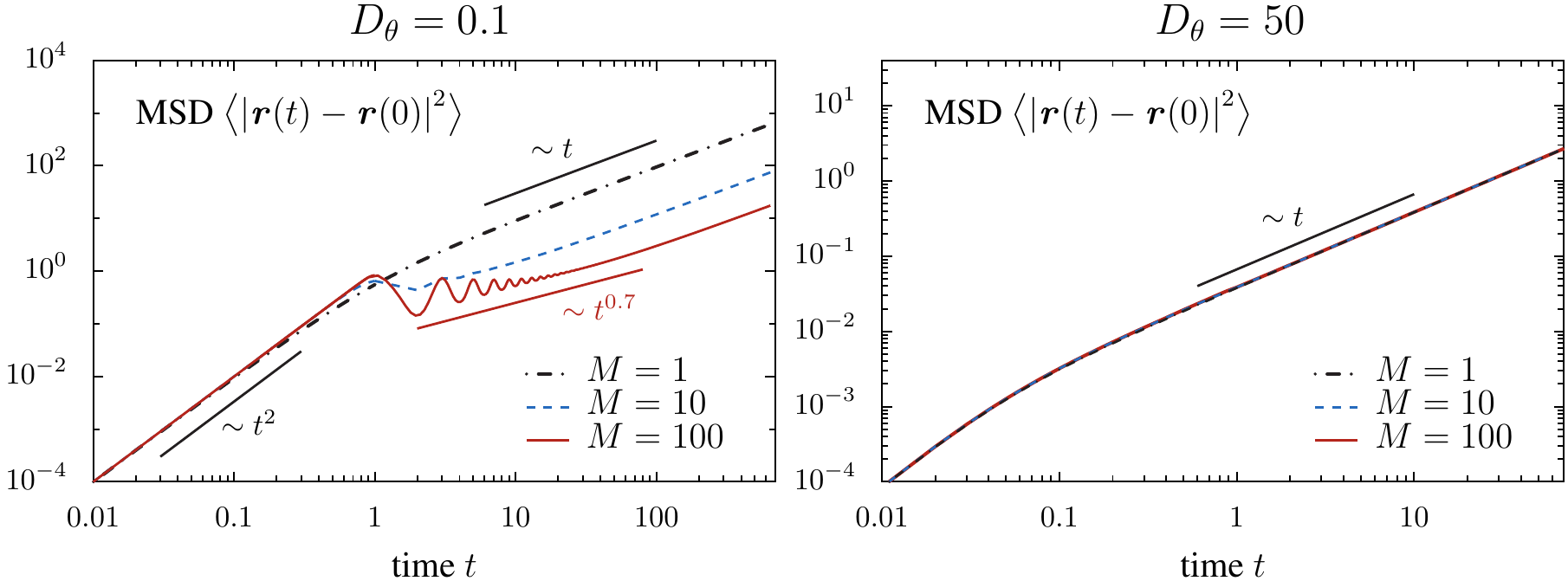}   
  \end{center}
  \caption{Mean squared displacement for several values of~$M$:~low noise regime~(\textit{left}) and high 
noise
(\textit{right}). In the low noise regime ($D_\theta < \lambda_r$), the mean squared displacement~(MSD) shows 
oscillatory behavior
for large $M$. In contrast, the reversal dynamics does not influence the diffusion properties for $D_\theta > 
\lambda_r$~(see main
text for an explanation). Parameters were adjusted such that the reversal frequency $\lambda_r = 1$ in all 
cases:~$\kappa = M$, $v_0
= 1$. The straight red line on the left indicates intermediate sublinear scaling due to oscillatory behavior 
of the MSD which
could erroneously be interpreted as subdiffusion if oscillations are not properly resolved in an experiment. 
We emphasize,
however, that the MSD increases \textit{always linearly} in the long-time limit in our model. }
  \label{fig:MSD}
 \end{figure}

The existence of oscillatory velocity correlations is crucial for the properties of the mean squared 
displacement which can show
oscillatory behavior as well. 
Typical time dependencies of the mean squared displacement in different regimes are shown in 
Fig.~\ref{fig:MSD}. 
The knowledge about the existence of (weakly) oscillating mean squared displacements is important for the 
analysis of experimental data which are not expected to show oscillations as clean as the theoretical results discussed here. 
As a result, the mean squared displacement may misleadingly suggest a subdiffusive regime due to the visual 
impression from noisy data (see red line on the left of Fig.~\ref{fig:MSD}). 
We emphasize, however, that our model does not predict subdiffusion but normal diffusion in the long-time 
limit. 

Since normal diffusion is expected, the motion is properly characterized by the spatial diffusion coefficient 
$\mathcal{D}$. 
Due to the preparatory work done in previous sections, its derivation is straightforward. 
The diffusion coefficient is obtained from the general solution, Eq.~\eqref{eqn1:diff:coeff}, by inserting the 
Laplace transform
of the correlation function of the reversal dynamics, Eq.~(\ref{eqn:red:corr:func:4}\textit{b}). 
For the clock model, we obtain the diffusion coefficient
\begin{eqnarray}
 \label{eqn:diff:coeff:spec:clock}
 \mathcal{D} = \frac{v_0^2}{2 D_\theta} \! \cdot \! \left [ 1 - \frac{2 \kappa}{D_\theta M} \! \cdot \! 
\frac{(\kappa +
D_\theta)^M - \kappa^M}{(\kappa + D_\theta)^M + \kappa^M} \right ] \! .
\end{eqnarray}

\begin{figure}[t]
\begin{center}
 \includegraphics[width=0.7\columnwidth]{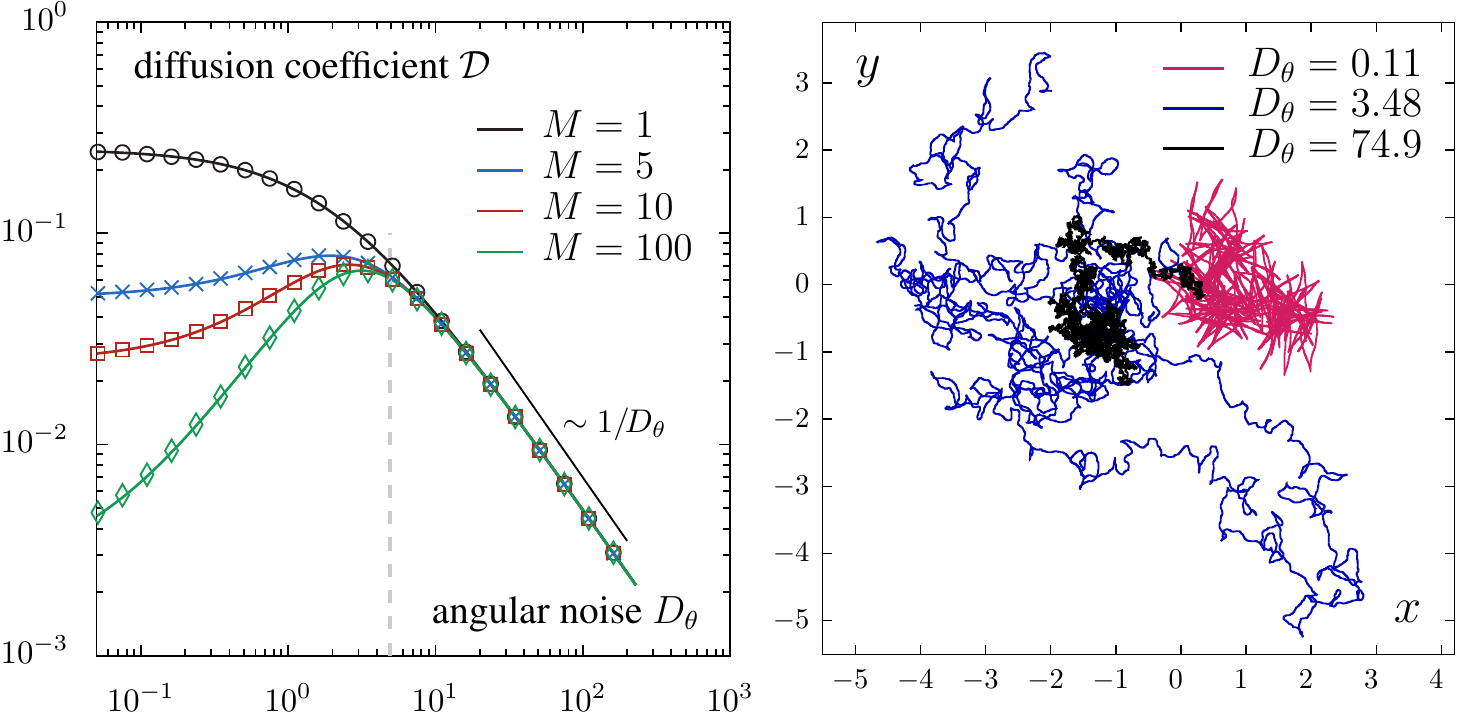}
\end{center}
\caption{\textit{Left}: Diffusion coefficient $\mathcal{D}$ as a function of the angular noise intensity 
$D_\theta$ from theory
\eqref{eqn:diff:coeff:spec:clock} shown by lines as well as numerical Langevin simulation (symbols). The light 
gray (dashed)
line indicates the estimate for the optimal noise strength from a comparison of relevant
timescales~(cf.~section~\ref{subsec:resonance}). \textit{Right}: Trajectories for three noise intensities 
$D_\theta$. The
diffusivity is maximal for an intermediate value $D_\theta$~(blue trajectory). Parameters: $v_0 = 1$, $\kappa 
= M$, $\lambda_r =
1$, numerical time step $\Delta t = 10^{-3}$. On the right: $M = 100$.}
\label{fig:D:noise_trajectories}
\end{figure}

In Fig.~\ref{fig:D:noise_trajectories}, the diffusion coefficient as a function of the angular noise is shown 
for several values
of $M$. 
The analysis reveals that a finite, optimal noise value $D_\theta$ exists which maximizes the diffusion 
coefficient. 
The existence of this maximum can intuitively be understood by looking at trajectories of particles in the 
different regimes, as shown in Fig.~\ref{fig:D:noise_trajectories} on the right. 
For low noise ($D_\theta \ll \lambda_r$), the particle moves back and forth along a line due to reversal. 
If, in addition, reversals occur in a fairly regular fashion, the trajectories are typically rather localized 
and the diffusivity
is low.  
On the other hand, the diffusion coefficient is likewise low for high angular noise ($D_\theta \gg \lambda_r$) 
since particles
perform a lot of turns thus preventing the departure from the initial position. 
Hence, the diffusivity is maximal for intermediate amplitudes of the rotational noise.  
The physics of this resonance effect is addressed in detail in the following section.

\section{Generalization -- reversal events as renewal process}
\label{sec:renewal}

In the previous section, we introduced an extension of a self-propelled particle model by 
including recurrent reversals of the direction of motion.
The internal particle dynamics was modeled by a clock representing the activation of certain biochemical 
processes that in turn trigger reversals events. 
Since the \textit{ticking} of the clock was assumed to be a stochastic process, subsequent reversals occur after 
stochastic waiting times. 
By construction, these waiting times are independent and identically distributed according to the run-time 
distribution $\varphi(t)$. 
Mathematically, this constitutes the definition of a renewal process~\cite{feller_an2_2008}~--~more precisely, 
the transition times determining the reversal dynamics $h(t)$ are controlled by a renewal process uniquely 
defined by a waiting time distribution $\varphi(t)$. 

In this section, we discuss general diffusion properties of self-propelled particles with directional reversal 
making use of the analogy to renewal theory. 
We relax the assumption that the run-time distribution $\varphi(t)$ is determined by the clock model thus 
considering arbitrary run-time distributions which may either be derived from more detailed models or measured 
experimentally. 

Previously, it was argued that the central object of interest is the correlation function of the reversal
process:~$C_{hh}(t,\tau) = \mean{h(t) h(t+\tau)}$.
This correlation function~$C_{hh}(t,\tau)$ can be determined in Laplace domain without specification
of the run-time distribution by using properties of renewal processes~(see~\cite{godreche_statistics_2001};~the derivation is
sketched in~\ref{sec:apA}).  
The derivation is based on similar ideas as the calculation presented in the context of the
clock model~(section~\ref{subsec:anaCM}). 
It turns out that Eq.~\eqref{eqn:gen:corr:func:laplace2}, 
 \begin{eqnarray}
  \label{eqn:gen:corr:func:laplace22}
   \widehat{\widehat{C}\,}_{\!\!hh}(s,u) = \frac{1}{u} \!\cdot\! \left [ \frac{1}{s} - \frac{2}{u-s} \!\cdot\!
\frac{\widehat{\varphi}(s) -
\widehat{\varphi}(u)}{ \left [1 + \widehat{\varphi}(u) \right ] \! \cdot \! \left [1 - \widehat{\varphi}(s) 
\right ] } \, \right
] \! , \! \!
 \end{eqnarray}
as well as the corresponding function in the long-time limit, Eq.~\eqref{eqn:red:corr:func:4},
 \begin{eqnarray}
 \label{eqn:red:corr:func:134}
   \widehat{C}_{hh}^{(eq)}(u) = 
  \frac{1}{u} \! \cdot \! \left [ 1 - \frac{2 \lambda_r}{u} \! \cdot \! \frac{1 - \widehat{\varphi}(u)}{1 +
\widehat{\varphi}(u)} \, \right ] \! , 
 \end{eqnarray}
constitute the correlation function for arbitrary run-time distributions. 
Accordingly, the Laplace transform $\widehat{\varphi}(s)$ of the run-time distribution~$\varphi(t)$ 
determines the correlation function of the reversal process. 
The inverse Laplace transformation in both arguments can be done analytically in special cases only. 
Note, however, that this transformation is not needed for the calculation of the diffusion coefficient
via Eq.~\eqref{eqn2:diff:coeff}. 

\subsection{Diffusion coefficient}
\label{subsec:gen:diffcoeff}

In section~\ref{sec:gen:dif:prop}, we derived a simple formula for the diffusion coefficient:~it is 
straightforwardly obtained from the Laplace transform of the correlation function~$C_{hh}^{(eq)}(\tau)$, 
Eq.~\eqref{eqn:red:corr:func:134}, by replacing the
variable $u$ by the noise amplitude $D_\theta$: 
  \begin{eqnarray}
    \label{eqn:D:GenSol}
  \mathcal{D} = \frac{v_0^2}{2 D_\theta} \cdot  \left [ 1 - \frac{2\lambda_r}{D_\theta} \cdot \frac{1 -
\widehat{\varphi}(D_\theta)}{1 + \widehat{\varphi}(D_\theta)} \right ] \! .
  \end{eqnarray}
This solution determines the diffusion coefficient for any run-time distribution $\varphi(t)$. 
Once $\varphi(t)$ has been derived from theoretical considerations~--~as it was done in the context of the 
clock model~--~or it
was fitted to experimental data, the diffusion coefficient can immediately be calculated.

In the following, we discuss the properties of this solution.   
First, we note that the diffusion coefficient of a self-propelled particle with reversal is always lower 
compared to a particle
which does never reverse its direction of motion if the trajectories are comparably persistent, i.e.~if 
$D_\theta$ is equal in
both cases. 
This can be seen from the fact that the term in brackets is always smaller than one since 
$\widehat{\varphi}(D_\theta) \in (0,1)$
for all $D_\theta > 0$: 
 \begin{eqnarray}
  \mathcal{D} = \frac{v_0^2}{2 D_\theta} \cdot  \left [ 1 - \frac{2\lambda_r}{D_\theta} \cdot \frac{1 -
\widehat{\varphi}(D_\theta)}{1 + \widehat{\varphi}(D_\theta)} \, \right ] <
\frac{v_0^2}{2 D_\theta} .
 \end{eqnarray}

Henceforth, we discuss two limiting cases, namely the high noise limit ($D_\theta \gg \lambda_r$) and the low 
noise
regime~($D_\theta \ll \lambda_r$). In the former case, we exploit that the Laplace transform of the waiting 
time
distribution tends to zero for large values of its argument. Therefore, we obtain
  \begin{eqnarray}
    \label{eqn:diff:coeff:gen:sol:gen:HN}
    \mathcal{D} \sim \frac{v_0^{2}}{2 D_\theta} \cdot \left [ 1 - \mathcal{O} \! \left ( \! 
\frac{\lambda_r}{D_\theta}
\right ) \right ] \! .
  \end{eqnarray}  
Hence, the diffusion coefficient coincides with the diffusion coefficient of non-reversing self-propelled
particles~\cite{schienbein_langevin_1993,mikhailov_self_1997}. 
This is plausible since reversals are not expected to influence the diffusion properties in the high noise 
regime as argued before in the context of the velocity correlation function, 
cf.~section~\ref{sec:subsec:res}.

We derive the low noise limit by expanding $\widehat{\varphi}(D_\theta)$ in a Taylor series: 
  \begin{eqnarray}
  \label{eqn:ser:psis:alMom}  
    \widehat{\varphi}(D_\theta) 
     &= \sum_{n=0}^{\infty} \frac{(-1)^n D_\theta^n}{n!} \int_0^{\infty} \! \rmd t \, t^n \, \varphi(t) . 
  \end{eqnarray}  
The series coefficients are determined by the moments\footnote{Here, we
assume that the moments exist and are finite.} of the waiting time distribution~$\varphi(t)$. 
However, it is more insightful to work with the central moments $\mean{\left ( \Delta t\right)^n} = 
\int_0^\infty \! \rmd t \, \left ( t - \mean{t} \right)^n \! \varphi(t)$. 
The diffusion coefficient is obtained by inserting this series into Eq.~\eqref{eqn:D:GenSol} and expanding the
resulting expression in powers of the noise:
  \begin{eqnarray}
    \label{eqn:diff:coeff:gen:sol:gen:LN2}
    \mathcal{D} = \frac{\lambda_r v_0^2}{4} \left [ \mean{\left ( \Delta t\right)^2} + 
     \frac{1 - 2\lambda_r^3\mean{\left ( \Delta t\right)^3}}{6} \cdot \frac{D_\theta}{\lambda_r^3} 
    + \mathcal{O} \! \left ( \frac{D_\theta^2}{\lambda_r^4} \, \right) \! \right ] \! .
  \end{eqnarray}
Interestingly, the diffusion coefficient is determined by the variance of the waiting time
distribution for low noise values. 
Thus, the diffusion coefficient tends to zero for small noise amplitudes, if the
reversal time distribution is narrow. Moreover, if the first Taylor coefficient is positive, i.e.
 \begin{eqnarray}
    \label{eqn:diff:coeff:max:cond}
     2 \lambda_r^3 \mean{\left ( \Delta t\right)^3} < 1 ,
 \end{eqnarray}
the diffusion coefficient \textit{increases} proportional to the noise strength. 
Since the dependence of the diffusion coefficient on the noise is continuous and the diffusion coefficient 
decreases for large noise values, there must exist a maximum in between: an optimal angular noise value 
maximizes the diffusivity. 
Eq.~\eqref{eqn:diff:coeff:max:cond} constitutes a sufficient condition for the existence of a maximum.

\subsection{Resonance - optimal noise maximizes diffusion}
\label{subsec:resonance}

 \begin{figure}[b]
  \begin{center}
  \includegraphics[width=0.55\columnwidth]{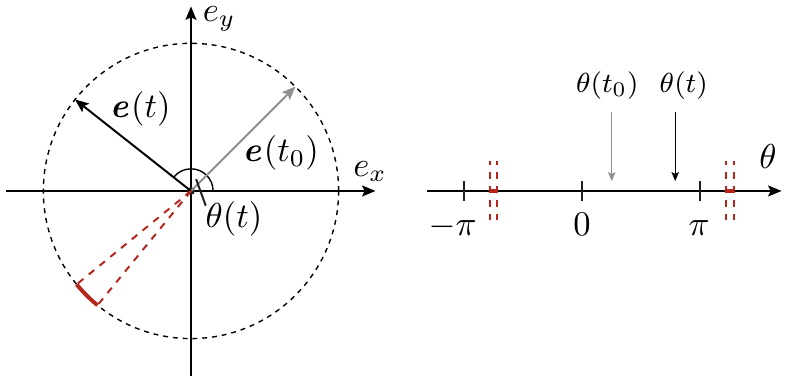}   
  \end{center}
  \caption{\textit{Left}:~Sketch of the temporal dynamics of $\vec{e}(t)$ determining the direction of motion 
of a particle. The
red cone indicates a region where $\vec{e}(t)$ points in the opposite direction with respect to the initial
direction of motion $\vec{e}(t_0)$. \textit{Right}:~The angles $\theta$ corresponding to the vectors shown on 
the left. }
  \label{fig:DirecScheme}
 \end{figure}

\begin{figure}[t]
  \begin{center}
  \includegraphics[width=0.5\columnwidth]{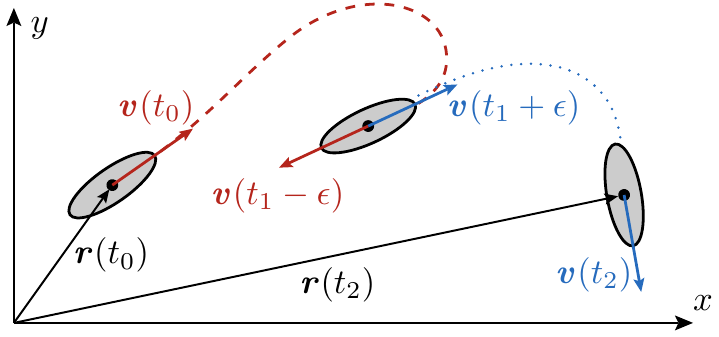}   
  \end{center}
  \caption{Illustration of the motion strategy for a maximal diffusion coefficient. A particle, initially 
located at
$\vec{r}(t_0)$, reverses its direction at time $t_1$. If the velocity vector prior to the reversal event 
$\vec{v}(t_1 -
\epsilon)$, where $\epsilon \ll 1$, is antiparallel to the velocity at time $t_0$, the reversal will increase 
the probability
that the particle departs from its initial position instead of moving back to the origin thus enhancing the 
diffusivity. }
  \label{fig:RevPartScheme}
 \end{figure}

The resonance effect -- the maximization of the diffusion coefficient for a finite angular noise intensity --
is understood by comparing the relevant timescales. 
For an illustration of the following arguments, see Fig.~\ref{fig:DirecScheme} and 
Fig.~\ref{fig:RevPartScheme}.

We give an intuitive argument why this maximum exists and estimate its position by rephrasing the problem as 
follows: 
\textit{Suppose, a particle reverses its direction of motion every $\lambda_r^{-1}$. How large should 
the angular noise strength $D_\theta$ be in order to maximize the diffusivity?}
To answer this question, we note first that a self-propelled particle moves ballistically into the direction 
determined by $\vec{e}(t)$, i.e.~it moves ballistically away from its initial position at small
timescales. 
However, angular fluctuations cause the orientation of the body axis $\vec{e}(t)$ to rotate. 
Apparently, the particle tends to move back to its initial position if $\vec{e}(t)$ points into the opposite 
direction with respect to $\vec{e}(t_0)$. 
Thus, a particle can move further away from its initial position, if it reverses its direction of
motion in this very moment. 
Hence, two relevant timescales exist:~(i)~the mean time between two reversal events and (ii)~the 
characteristic time it takes for the body axis to rotate by approximately $180^\circ$ driven by rotational 
noise.
The former is determined by the inverse reversal frequency $\lambda_r^{-1}$. 
In order to estimate the latter, we note that the dynamics of the angle $\theta$ is equivalent to Brownian 
motion with diffusion coefficient $D_\theta$ in one dimension thus constituting a mean first-passage time 
problem:~\textit{What is the mean time $\tilde{\tau}$ it takes for a Brownian particle with diffusivity 
$D_\theta$ to escape out of the interval $(\theta_0-\pi,\theta_0+\pi)$, given that the initial position was 
$\theta_0$?}
The solution of the first-passage time problem yields the well known diffusion law:~$\pi^2 = 2 D_\theta 
\tilde{\tau}$. 
The comparison of both timescales, $\tilde{\tau} \overset{!}{=} \lambda_r^{-1}$, may be used to estimate the 
optimal noise value: 
 \begin{eqnarray}
   \bar{D}_\theta \approx \frac{\, \lambda_r \!\, \pi^2}{2} \, .   
 \end{eqnarray}
This reasoning is valid if the first-passage time distribution as well as the reversal time distribution are
sufficiently narrow. 
However, we obtain a rather reasonable estimate for the optimal noise
amplitude as shown by the gray dashed line in Fig.~\ref{fig:D:noise_trajectories}. 

\section{Summary \& outlook}
\label{sec:summary}

In this work, we studied the diffusion properties of self-propelled particles that repeatedly reverse
their direction of motion.
We adopted a coarse-grained viewpoint aiming at describing the reversal dynamics phenomenologically and 
discussing the effects of the directional reversal on the diffusion properties of active particles within the 
framework of stochastic processes. 
For this purpose, we model individual particles as point-like objects with a propelling engine allowing for 
active motion at constant speed. 
Fluctuations of the driving motor as well as external heterogeneities are taken into account by 
addition of rotational noise. 
The internal dynamics of the propelling engine that controls reversal events is modeled by a
simple clock model, where the ticks of the clock represent biochemical activation processes. 
We derived results for velocity correlation functions, mean squared displacement and, in particular, the 
diffusion coefficient for this model. 
Notably, we found that the mean squared displacement can show oscillatory behavior for intermediate times. 
Therefore, experimental data must be analyzed carefully because the visual impression of noisy data 
may wrongly be interpreted as a subdiffusive regime if oscillations are not properly resolved. 

In the second part, we generalized the results obtained from the clock model describing the reversal dynamics 
as a renewal process:~subsequent reversal events occur after random waiting times which are distributed 
according to a given run-time distribution. 
Given a run-time distribution, we derived a general formula for the diffusion coefficient. 
Our analysis reveals that an optimal rotational noise value maximizes the diffusivity if the run-time 
distribution is sufficiently narrow. 
This resonance effect can be understood as a matching of timescales of the rotational diffusion and
the mean time between two reversals. 

\begin{figure}[t]
 \begin{center}
   \includegraphics[width=0.85\textwidth]{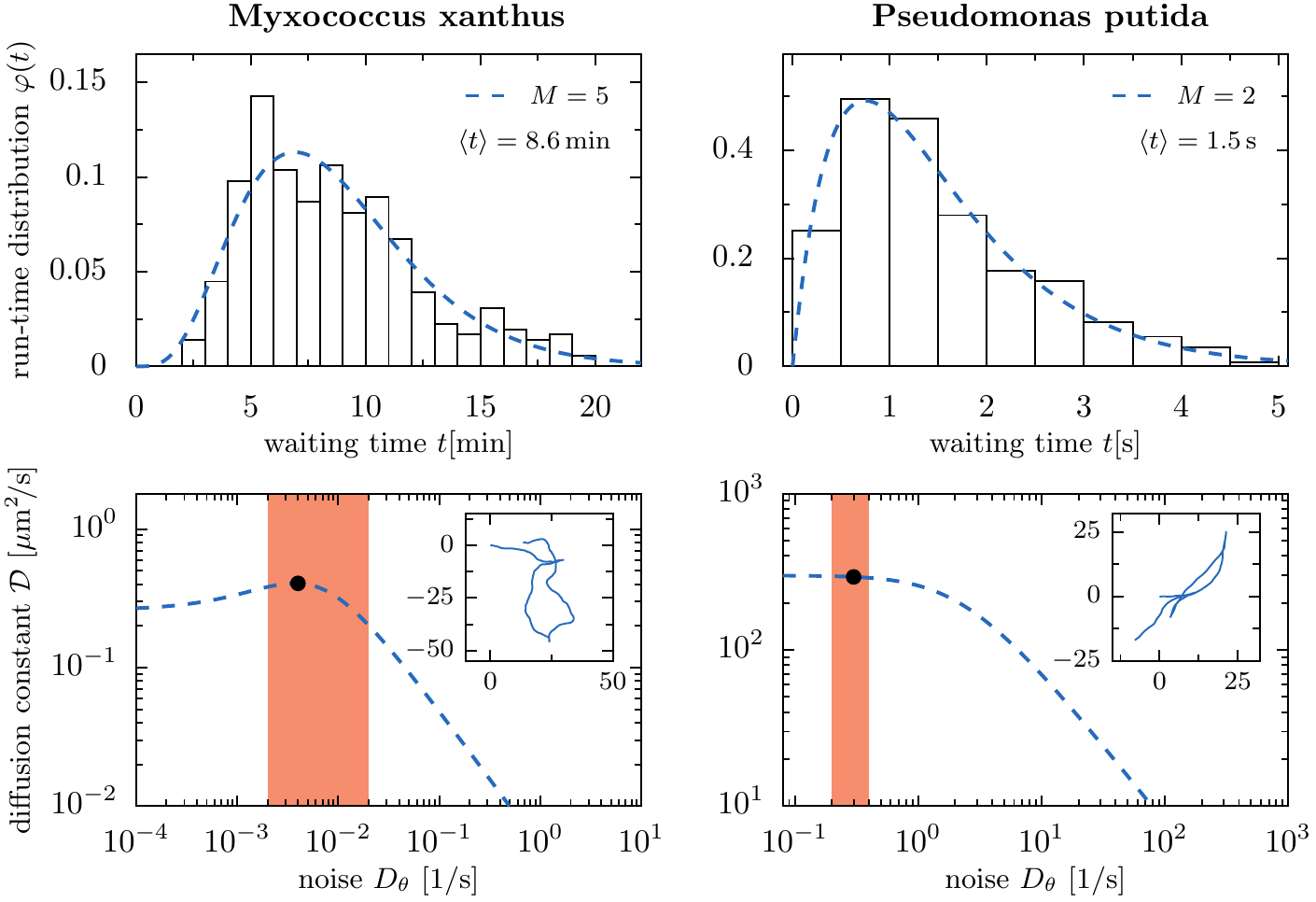}   
 \end{center}
 \caption{\textit{Upper panel}:~run-time distribution~$\varphi(t)$ for the bacteria \textit{Myxococcus 
xanthus}~(left) and \textit{Pseudomonas putida}~(right). 
Experimentally observed distributions of myxobacteria and \textit{Pseudomonas putida}~--~reproduced 
from~\cite{Wu_periodic_2009} and~\cite{theves_bacterial_2013}, respectively~--~are represented by bars in the 
two upper panels. The lines represent the corresponding waiting time distributions from the clock model, 
Eq.~\eqref{eqn:WTPDF:Clock}, where parameters were estimated from the data according to 
Eq.~\eqref{eqn:MapExpTheo}. \newline
\textit{Lower panel}:~diffusion coefficient~$\mathcal{D}$ as predicted by the clock model, 
Eq.~\eqref{eqn:diff:coeff:spec:clock}. 
We used the following estimates for the mean time in between two reversals obtained from the distributions 
above:~$\mean{t} = 8.6\,$min and $\mean{t}=1.5\,$s, respectively.  
A characteristic speed was estimated to be $v_0 \approx 0.1\,\mu$m$/$s for myxobacteria~\cite{sliusarenko_aggregation_2007} and 
$v_0 \approx 40\,\mu$m$/$s for \textit{Pseudomonas putida}~\cite{theves_random_2015}. 
The insets show a characteristic trajectory (spatial scale in $\mu{m}$) for the noise amplitude indicated by a black dot 
in the main figure. The trajectories represent a time windows of $\Delta t = 26\,$min (left) and $\Delta t = 3\,$s (right). 
The shaded regions (orange) indicate noise amplitudes expected for myxobacteria, $D_\theta \in 
(0.002,0.02)\,$s$^{-1}$~\cite{Jelsbak_pattern_2002,sliusarenko_aggregation_2007}, and \textit{Pseudomonas 
putida}, $D_\theta \in (0.2,0.4)\,$s$^{-1}$~\cite{theves_random_2015}, respectively. }
 \label{fig:linkExp}
\end{figure}

We conclude by discussing the potential relevance of this resonance effect in microbiological systems by 
estimating and comparing the order of magnitude of relevant time and length scales from experimental data 
obtained in previous 
works~\cite{Wu_periodic_2009,theves_bacterial_2013,sliusarenko_aggregation_2007,theves_random_2015, 
Jelsbak_pattern_2002}. 
As an example, we  consider two bacterial species:~\textit{Myxococcus xanthus} and \textit{Pseudomonas putida}\footnote{We note 
that \textit{Pseudomonas putida} was shown to exhibit a more complex motion pattern than considered in this work:~forward and 
backward motion occur with different speeds~\cite{theves_bacterial_2013}. Here, we do not intend to describe these bacteria in 
full detail but use the illustrative example to estimate the order of magnitude of characteristic quantities 
such as the diffusion coefficient.}, both showing directional 
reversals~\cite{Jelsbak_pattern_2002,sliusarenko_aggregation_2007,Wu_periodic_2009,theves_bacterial_2013,theves_random_2015}. 
The results are summarized in Fig.~\ref{fig:linkExp}. 
We first estimated the coefficients~$\{M,\kappa\}$ from the experimentally observed run-time distributions~$\varphi(t)$ via 
the relations of these model parameters to the mean $\mean{t}$ and the variance $\sigma^2 = 
\mean{t^2}-\mean{t}^2$ of the run-time distribtion~(cf.~section~\ref{subsec:IntoCM}): 
 \begin{eqnarray}
   \label{eqn:MapExpTheo}
   \kappa = \frac{\mean{t}}{\sigma^2} \, , \quad M = \frac{\mean{t}^2}{\sigma^2} \, . 
 \end{eqnarray}
Besides the characteristics of the reversal process, we estimated characteristic speeds:~$v_0 \approx 0.1\,\mu$m$/$s for 
myxobacteria~\cite{sliusarenko_aggregation_2007} and $v_0 \approx 40\,\mu$m$/$s for \textit{Pseudomonas 
putida}~\cite{theves_random_2015}. 
Knowing the characteristic speed $v_0$, the order of magnitude of the rotational noise~$D_\theta$ can be estimated from 
the persistence of trajectories as provided in~\cite{Jelsbak_pattern_2002}, or from direct measurements of the velocity 
correlation function~\cite{theves_random_2015}. 
However, we expect the persistence of trajectories and, consequently, the angular noise to be highly dependent on the 
environmental conditions and therefore do only consider rough estimates summarized in Tab.~\ref{tab:estimates}.

\begin{table}[t]
\caption{Summary of estimated characteristic parameter values. }
\label{tab:estimates}
\begin{indented}
\item[]\begin{tabular}{@{}ll}
\br
\textit{Myxococcus xanthus} & \textit{Pseudomonas putida} \\
\mr
 $v_0 \approx 0.1\,\mu$m$/$s~\cite{sliusarenko_aggregation_2007} & $v_0 \approx 40\,\mu$m$/$s~\cite{theves_random_2015} \\
 $D_\theta \in (0.002,0.02)\,$s$^{-1}$~\cite{Jelsbak_pattern_2002,sliusarenko_aggregation_2007} & $D_\theta \in 
(0.2,0.4)\,$s$^{-1}$~\cite{theves_random_2015} \\
 $\mean{t} \approx 8.6\,$min~\cite{Wu_periodic_2009} & $\mean{t} \approx 1.5\,$s~\cite{theves_bacterial_2013}\\
\br
\end{tabular}
\end{indented}
\end{table}

We find that the clock model provides an excellent fit to the run-time distributions for myxobacteria and \textit{Pseudomonas 
putida}, even though the characteristic time scales of the two species differ by one order of magnitude, as illustrated in
Fig.~\ref{fig:linkExp}. 
Apparently, the reversal processes are not Poisson processes, since we obtain $M > 1$ in both cases. 
Furthermore, we find that an optimal rotational noise amplitude can exist indeed in the case of myxobacteria, which is not the 
case for \textit{Pseudomonas putida}. 
The shape of  trajectories in Fig.~\ref{fig:linkExp} suggest that the diffusion of \textit{Pseudomonas putida} is dominated by reversal events 
whereas, in contrast, the timescales of rotational diffusion and reversal coincide roughly for myxobacteria.
Our analysis suggests that the coincidence of rotational noise and reversal frequency leads to an optimal 
(maximal) diffusion. 
It will be very interesting to check experimentally whether the \textit{natural parameters} of other microbiological systems 
were evolutionary tuned in such a way that microorganisms are best adapted to their environment in the sense that their 
diffusivity is optimal~--~a prerequisite for an optimal food search strategy. 

A further aspect where the described model for reversing bacteria may become important is the modeling of 
collective behaviors like rippling, clustering and other forms of collective motion of bacteria. 
Earlier studies on myxobacterial rippling~\cite{sliusarenko_accordion_2006,boerner_generalized_2006} compared 
simulation results with experimental data~\cite{sliusarenko_accordion_2006,Welch_cell_2001} based on the reversal 
time statistics of labeled bacteria in colonies assuming identical cell behaviors. 
The presented model, in contrast, allows for the modeling of the variability of individual cells that is 
expressed in the run-time distributions displayed in Fig.~\ref{fig:linkExp}.

	
%
\appendix

\section{Reversal as a renewal process}
\label{sec:apA}

In this section, we sketch the calculation which reveals that the probabilities $P_k(t,\tau)$ to observe
exactly $k$ reversal events in the time interval $[t,t+\tau]$ are determined by
\eqnlabel{eqnA:jumps}
\Anumparts
\begin{eqnarray}
 \!\!\!\widehat{\widehat{P}\,}_{\!\!0}(s,u) &= \frac{1}{u} \! \cdot \! \left [ \frac{1}{s} - \frac{1}{1 - 
\widehat{\varphi}(s)}
 \! \cdot \! \frac{\widehat{\varphi}(s) -
\widehat{\varphi}(u)}{u - s} \right
] \! ,\\
 \!\!\!\widehat{\widehat{P}\,}_{\!\!k}(s,u) &= \frac{1 - \widehat{\varphi}(u)}{u} \! \cdot \! \frac{\left [ 
\widehat{\varphi}(u)
\right ]^{k-1}}{1 - \widehat{\varphi}(s)} \! \cdot \! \frac{\widehat{\varphi}(s) - \widehat{\varphi}(u)}{u - 
s} 
\end{eqnarray}
\endAnumparts
in Laplace domain for arbitrary waiting time distributions $\varphi(t)$, cf.~\eref{eqn:probab:jumps}. Accordingly, the correlation function
$\widehat{\widehat{C}\,}_{\!\!hh}(s,u)$, cf.~\eref{eqn:gen:corr:func:laplace2}, which was derived in section~\ref{sec:ClockModel} for the
clock model, is valid for arbitrary waiting time distributions as well. The derivation is based on standard properties of renewal
processes~\cite{godreche_statistics_2001,klafter_first_2011}. 

At first, we assume that the observation of a particle is started at $t=0$. The probability density function $\varphi_1(t)$ for the
occurrence of one reversal event is equal to the run-time distribution $\varphi(t)$. The probability density function $\varphi_k(t)$ for the
occurrence of $k$ reversal events is determined by multiple convolutions of the run-time distribution with itself: 
\Anumparts
 \begin{eqnarray}
   \varphi_2(t) &= \int_0^t \rmd t' \varphi(t') \varphi(t-t') = \varphi(t) * \varphi(t)\\
   \varphi_3(t) &= \int_0^t \rmd t' \varphi(t') \varphi_2(t-t') = \varphi(t) * \varphi_2(t) = \varphi(t)*\varphi(t)*\varphi(t)\\
   & \;\,\vdots \nonumber \\
   \varphi_k(t) &= \underbrace{\varphi(t) * \varphi(t) * ... * \varphi(t)}_{\mbox{\normalfont $k$-fold}} .
 \end{eqnarray}
\endAnumparts
According to the convolution theorem of the Laplace transform, a convolution is reduced to a multiplication in the Laplace
domain:  
 \begin{eqnarray}
  \widehat{\varphi}_k(s) = \left [\textcolor{white}{\frac{a}{a}} \!\!\!\! \widehat{\varphi}(s) \right ]^{k} \! . 
 \end{eqnarray}

Now, we consider the situation where the observation is started at an arbitrary time~$t$. The observation will surely begin in between two
reversal events:~the particle has reversed its direction of motion a certain number of times before time $t$, and will reverse again after
the waiting time $\tau_{+}$ measured from the beginning of the observation (\textit{forward waiting time}). The statistics of $\tau_{+}$ is
different from the usual run-time distribution because the measurement was started between two reversals. However, the probability density
for $\tau_+$ can be expressed by the run-time distribution as follows: 
 \begin{eqnarray}
  \label{eqn:FWTPDF}
  \Phi_1(t;\tau_+) = \sum_{k=0}^\infty \int_0^{t} \! \rmd t' \, \varphi_k(t') \varphi(t + \tau_+ - t') . 
 \end{eqnarray}
The integrand reflects the probability that $k$ reversals occurred up to time~$t'$ and the next reversal is observed at time $t+\tau_+$.
However, neither $t'$ nor the number of reversals $k$ is known and, therefore, one has to integrate and sum over these quantities,
respectively. 

Equation \eref{eqn:FWTPDF} is rather difficult to handle. In contrast, its Laplace transform takes a simple form. The transformation is
performed in both arguments, where $s$ is conjugate to $t$ and $u$ is conjugate to $\tau_+$:
 \begin{eqnarray}
  \widehat{\widehat{\Phi}}_1(s;u) = \frac{1}{1 - \widehat{\varphi}(s)} \frac{\widehat{\varphi}(s) - \widehat{\varphi}(u)}{u-s} .
 \end{eqnarray}
The statistics of the second and subsequent reversals follows the usual run-time distribution. 

The previous considerations allow the straightforward calculation of the probabilities $P_k(t,\tau)$. The probability not to reverse is
determined by the probability not to observe the first jump within the observation time $\tau$: 
 \begin{eqnarray}
   P_0(t,\tau) = 1 - \int_0^{\tau} \rmd \tau_+ \, \Phi_1(t;\tau_+). 
 \end{eqnarray}
The probability to observe exactly one reversal event, $P_1(t,\tau)$, is determined by the probability to observe the first reversal event
within the observation time and no subsequent reversals: 
 \begin{eqnarray}
   \label{eqn:chi1}
   P_1(t,\tau) = \int_0^{\tau} \rmd \tau_+ \, \Phi_1(t;\tau_+) \Phi(\tau - \tau_+). 
 \end{eqnarray}
In this equation, the probability that a second reversal event does not occur was introduced (\textit{survival probability}):  
 \begin{eqnarray}
   \Phi(t) = 1 - \int_0^t \rmd t' \, \varphi(t') . 
 \end{eqnarray}
Further, $P_k(t,\tau)$ can be expressed in the following form: 
 \begin{eqnarray}
   P_k(t,\tau) = \int_0^{\tau} \rmd t' \left [ \int_0^{t'} \rmd t'' \, \Phi_1(t;t'') \varphi_{k-1}(t' - t'') \right ] \! \Phi(\tau -
t'). 
 \end{eqnarray}
The inner integral represents the probability that the first reversal event is followed by $k-1$ additional reversals. This expression is
convolved by the survival probability reflecting that a $(k+1)$th reversal is not observed within the observation time $\tau$. 

The Laplace transforms of the integral expressions for $P_k(t,\tau)$, which are obtained by multiple application of the convolution theorem,
finally yield the algebraic relations~\eref{eqnA:jumps}. 

\section*{Acknowledgement}

RG and MB gratefully acknowledge the support by the German Research Foundation via Research Training Group~1558. 
FP acknowledges support by the Agence nationale de la recherche via JCJC project \textit{BactPhys} as well as project ANR-15-CE30-0002-01
and by the F\'{e}d\'{e}ration W. D\"oblin (CNRS) via AxePhysBio project \textit{The Physics of Bacterial Invasion}. 

\section*{References} 


\input{SPP_Reversal.bbl}

\end{document}

%% file: SPP_Reversal.bbl
\providecommand{\newblock}{}